# Impact of Fe doping on the electronic structure of SrTiO$_3$ thin films determined by resonant photoemission

J. Szade[1,2], D. Kajewski[1], J. Kubacki[1,2], K. Szot[1,3], A. Köhl[3], Ch. Lenser[3], R. Dittmann[3]

[1] A Chełkowski Institute of Physics, University of Silesia, Katowice, Poland,

[2] Silesian Center for Education and Interdisciplinary Research, Chorzów, Poland

[2] Peter Grünberg Institut, Forschungszentrum Juelich, 52425 Juelich, Germany

Abstract

Epitaxial thin films of Fe doped SrTiO$_3$ have been studied by the use of resonant photoemission. This technique allowed to identify contributions of the Fe and Ti originating electronic states to the valence band. Two valence states of iron Fe$^{2+}$ and Fe$^{3+}$, detected on the base of XAS spectra, appeared to form quite different contributions to the valence band of SrTiO$_3$. The electronic states within the in-gap region can be attributed to Fe$^{2+}$, Fe$^{3+}$ and Ti$^{3+}$ ions. Fe$^{2+}$ originating states which can be connected to the presence of oxygen vacancies form a broad band reaching binding energies of about 0.5 eV below the conduction band while Fe$^{3+}$ states form in the gap a sharp feature localized just above the top of the valence band. It has been shown that Fe doping induced Ti originating states in the energy gap which can be related to hybridization of Ti and Fe 3d orbitals.

**Introduction**

SrTiO$_3$ (STO) can be regarded as a model oxide with the perovskite structure. Its band gap of 3.2 eV, a high dielectric constant and paraelectric state at the room temperature cause that STO is an interesting material for applications in emerging oxide electronics. Electrical conductivity of STO can be tuned by chemical doping, oxygen deficit, or formation of interfaces with other perovskites as it is for example for the LaAlO$_3$/STO systems [1]. Moreover, local insulator-metal transitions were observed at the nanoscale which was attributed to a strong coupling of the structural inhomogeneity and the modified electronic structure [2]. Doping and presence of defects or interfaces can lead to modifications of the energy gap region via the formation of new electronic states and transformation of the electronic wavefunctions from the localized to itinerant character. Therefore, knowledge on the electronic states responsible for such transformations is crucial to understand the electrical properties of a material which is insulating when undoped and structurally unchanged.

Slightly Fe doped STO is a model representative of acceptor doped large band-gap electroceramics [3, 4]. The mixed valence (Fe3+/Fe4+) character of the dopant creates a charge imbalance with respect to the lattice which is compensated by oxygen vacancies in a large oxygen partial pressure regime. [5, 6]. Full substitution of Ti by Fe leads to a metallic perovskite compound - $SrFeO_3$ which has a different structure than STO and is a helical antiferromagnet with $T_N$ = 134 K [7]. In the recent paper da Silva et al. have shown that synthesis of $Sr(Ti,Fe)O_3$ powders with nanometer-sized particles by the polymeric precursor method leads to formation of cubic perovskite structure in the whole substitution range [8,9]. They showed that $Sr(Ti,Fe)O_3$ is a mixed (ionic and electronic) conductor and that the electronic contribution becomes dominant with increasing Fe content. Other theoretical approaches to the Fe doping effects have shown a formation of the electronic in-gap states. Their degree of localization and bandwidth depend on the doping level, the assumed Fe valence and the method employed for the calculations [10-12].

For the majority of applications, the use of thin films is inevitable. Due to the strongly kinetically limited growth conditions in many widely used thin film deposition methods, e.g. pulsed laser deposition (PLD) and sputtering, the formation of defects and the incorporation of dopants is into the crystal lattice considerably differ from single crystals and bulk ceramics. Fe doped STO films exhibit, even when they are grown at high oxygen pressure, no significant $Fe^{4+}$ contribution [13]. Analysis of the Fe 2p photoemission multiplet in single crystalline STO thin films showed that for doping levels of Fe up to 2 % its valence state is rather a combination of 2+ and 3+ with a divalent contribution situated mostly on the surface [14]. XAS studies confirmed the mixed valence state of iron and have shown that it is independent of the doping concentration and that annealing increases the relative $Fe^{2+}$ signal ratio [14]. Our XPS results obtained for Fe doped STO films indicated to a clear relation between doping and the increased density of states in the energy gap [15The effect of doping was discussed in terms of coherent and incoherent electronic states mainly for electron doping as it was done for Nb replacing Ti in STO [16]. It was also shown that thermal reduction of undoped STO leads to the appearance of the metallic like bands crossing the Fermi level [17,18]. However, no detailed studies of the in-gap electronic structure for acceptor-like dopants such as Fe has been reported so far.

In this paper we aim to clarify the role of Fe in the modification of the STO electronic structure. By using the resonant photoemission method, we were able to obtain the partial densities of states for Fe, Ti and O. We show also that taking into account the XAS and photoemission data we are able to clearly distinguish the contribution of the divalent and trivalent iron ions to the electronic states in the valence band. We determined also electronic states in the energy gap which originate from Ti states and investigated how they how they depend on iron doping.

We compare our experimental data with calculations from the literature and analyse them in terms of electronic states localization.

**Experimental**

Thin films of $SrTiO_3$ (STO) undoped and doped with Fe were grown by PLD using a KrF excimer laser and ceramic targets with nominal composition $SrTiO_3$, $SrTi_{0.99}Fe_{0.01}O_3$, $SrTi_{0.98}Fe_{0.02}O_3$ and $SrTi_{0.95}Fe_{0.05}O_3$ respectively. The films had the thicknesses of about 20 nm. The substrates were commercially available (001) oriented conducting Nb:STO single crystals in order to prevent charging during spectroscopic investigations. The deposition process was performed in an oxygen partial pressure of 0.25 mbar at a repetition rate of 5 Hz and laser energy density of 0.8 J /cm2. The substrate was kept at temperature of 700°C. X-ray diffraction and electron diffraction (LEED) examination confirmed the epitaxial growth of the deposited films.

Photoelectron spectroscopy measurements and the X-ray absorption studies (XAS) were performed at the I311 beamline in Max-lab synchrotron. The X-ray beam size was of the order of 100x500 $\mu m^2$. The XAS spectra we present in this paper were obtained by the Auger electron yield (AEY) method based on intensity changes of a selected Auger electron peak.

The films were annealed prior to the measurements in UHV conditions at temperatures of 150°C, 300°C or 630°C. Heating caused removal of most carbon containing contaminations. Some remaining carbon and minor contribution of oxygen species which could not be attributed to perovskite structure could be detected for most samples. The effect of these contaminations was estimated and found to be irrelevant for the studies of the Fe electronic state.

**Results and discussion**

The resonant photoelectron spectroscopy (RESPE) is a technique which requires measurements of the photoelectron spectra for photon energies within the region of a selected absorption edge. Acquiring of the XAS spectrum is thus a first step for the RESPE studies.

Analysis of the XAS spectra obtained for the undoped STO film and the films with Fe content up to 5 % allowed to distinguish between $Fe^{2+}$ and $Fe^{3+}$ contribution to the observed spectra and to state that the XAS Fe L egde spectra have the same structure for all films. All spectra have two maxima within the $L_3$ edge and at least three components of the $L_2$ edge. Figure 1 presents the example of Fe $L_{2,3}$-edge XAS spectrum for the 2% Fe doped $SrTiO_3$ film measured in Auger Electron Yield mode after annealing the sample at 300°C for 30 min in ultra-high vacuum.

The observed maximum at about 708.8eV (marked as "A") comes mostly from the $Fe^{2+}$ contribution to the Fe $L_3$ absorption edge spectrum. The second observed maximum at 710.5 eV (marked as "B") can be attributed mostly to the $Fe^{3+}$ contribution [14]. However, one has to remember that the calculated structure of the XAS spectra shows some contributions from both valence states within the peaks A and B. A clear difference in the resonance behavior of the valence band, discussed below, for the photon energies which are very close to each other can confirm the different electronic configurations responsible for appearance of peaks A and B.

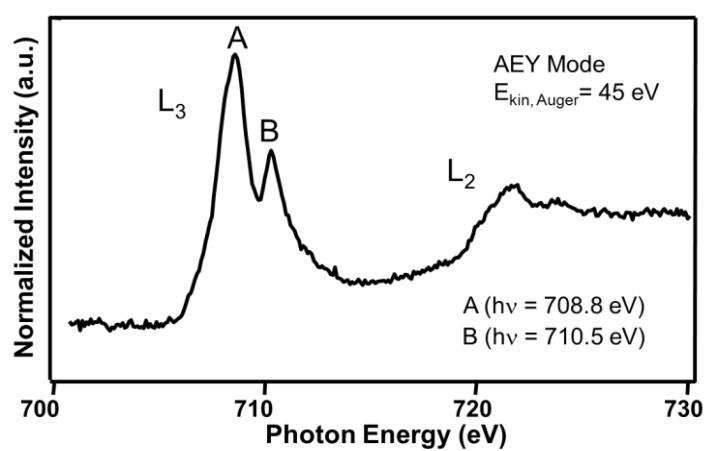

Fig. 1 Fe L edge x-ray absorption spectrum obtained in the Auger electron yield mode for the STO film doped with 2 % Fe. Peaks described as A and B can be attributed to mostly divalent and trivalent Fe ions according to the reasoning presented in the text.

The photoemission map of the valence band for the photon energy range of the Fe 2p-3d resonance is presented in Figure 2. The scale of intensity was tuned to highlight a subtle effects in the in-gap binding energy range. The main contribution is placed in the region of the valence band formed by the O, Sr and Ti orbitals, i.e. between 3 and 7 eV. This region was investigated many times for STO and the mentioned assignment of the electronic states is in agreement with the resonant photoemission studies of STO and calculations [see e.g. 9]. It is also noteworthy that some enhancement of photoemission is observed in the energy range over 8 eV where the electronic states are most likely related to defects [19].

In the photon energy range of the Fe $L_3$ absorption edge one could notice a clear resonant enhancement of the in-gap electronic states. The region of the Fe $L_2$ edge shows a very weak effect.

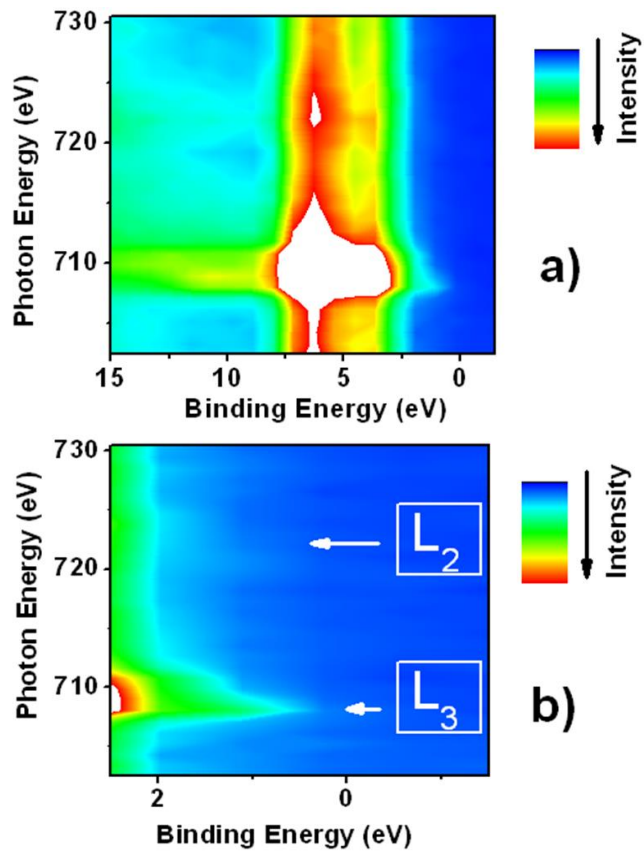

Fig. 2  Photoemission intensity maps in the valence band region obtained for the 2% Fe doped STO film for photon energies within the Fe 2p-3d resonance. Panel b) shows the zoomed region of the energy gap.

Surprisingly, there are almost no observable changes of the photoemission for the photon energy range of the $L_2$ edge. Such situation can be expected only when a high spin state of the Fe 3d level is realized and spin forbidden transitions are damping the resonance effect. Due to the spin exchange interaction and spin-orbit splitting the $L_2$ ($2p_{1/2}$) sublevel is spin polarized opposite to the majority spin of the 3d level. During the 2p-3d absorption process only the minority spin levels are virtually occupied. The resonant photoemission is then blocked from the occupied majority spin 3d subbands. In the high spin Fe $3d^5$ state it concerns both crystal field split sublevels $t_{2g}$ and $e_{2g}$. Having the XAS spectra for all investigated samples we could perform resonant photoemission spectroscopy of the valence band with the use of three different photon energies, to obtain partial (projected) density of states (PDOS) which, according to the discussion presented in our previous paper, correspond mostly to $Fe^{2+}$ and $Fe^{3+}$.

Two chosen photon energies were equal to the energies of the peaks A and B, and the third one was chosen for the off-resonance energy which was about 700 eV. In order to obtain PDOS of iron we subtracted spectra obtained for the on- and off-resonance energies after normalizing the spectra with the use of beam intensity. Fig. 3 shows three spectra obtained at energies of 700, 708.8 and 710.5 eV.

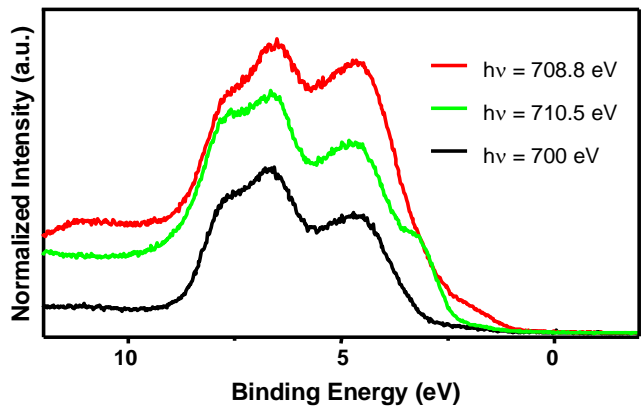

Fig. 3. Selected valence band photoemission spectra for 2% Fe doped STO film, obtained for photon energies on- and off-resonance. Photon energies of 708.8 eV and 710.5 eV correspond to the positions of the peaks in the XAS spectrum. 700 eV is the off-resonance photon energy

Calculated in such way, iron dependent DOS's for $Fe^{2+}$ and $Fe^{3+}$ are presented in Figure 4. Measurements were performed for three concentrations of iron in the films: 1, 2 and 5 at% and for two different ways of thermal treatment.

Several important features can be noticed – the structure of the Fe PDOS is quite different for two photon energies on resonance. It seems thus reasonable to attribute the obtained densities of states to $Fe^{2+}$ and $Fe^{3+}$. We will refer further in the text to the 3+ and 2+ iron valence states having in mind that this assignment is valid on the base of the XAS analysis and data from the literature [15].

The structure is different for the main part of the DOS i.e. for binding energies between 3-8 eV and within the band gap. For the $Fe^{2+}$ DOS the contribution to in-gap states is much more pronounced and extends to binding energies of about 0.5 eV – very close to the Fermi level. For the $Fe^{3+}$ the electronic states form a more complicated structure. A most interesting feature is a distinct peak situated at 2.5 eV, so just above the top of the valence band. Its evolution with the Fe concentration is also interesting. It is most pronounced for the lowest Fe concentration. It allows to relate that electronic state to acceptor like states coming from the isolated doping ions. The position of the peak agrees roughly with the data of ionization energy of Fe optical centers in STO [20]. Increasing doping level leads probably to more non-homogeneous Fe distribution and finally to formation of less localized electronic states. We found no effect described in the work of da Silva et al. and Rotschild et al [9,21] where the gradual decrease of the energy gap was detected with increasing Fe content at least in the concentration region studied in our films.

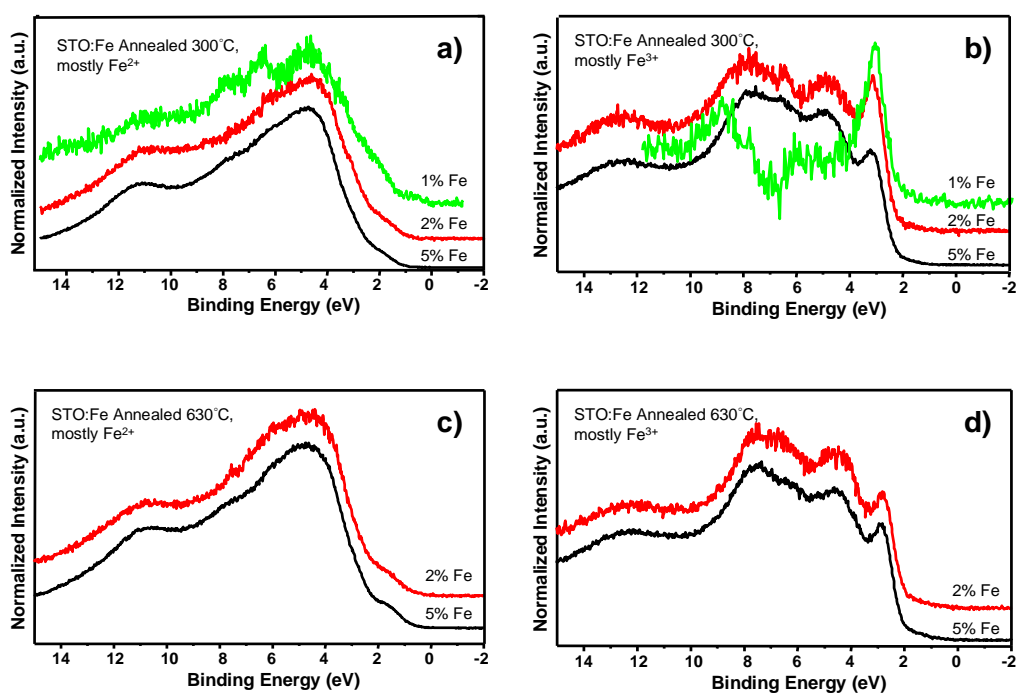

Fig. 4. Fe partial densities of states (PDOS) obtained for the resonance I corresponding to photon energy of 708.8 eV and peak A in the XAS spectra (a) and (c), and for photon energy of resonance II – peak B (710.5 eV) – panels (b) and (d). Panels (a) and (b) correspond to films annealed at 300°C prior to measurements while panels (c) and (d) were obtained for annealing at 630°C.

Fig. 5 shows the resonant behavior of the in-gap states presented in the CIS (constant initial state) mode for photon energy range of the Fe L absorption edge. The enhancement of the Fe originating states is well visible for binding energies starting from about 0.5 eV. No states at the Fermi level are observed what is in agreement with the classical description of acceptor states above 1 eV from the valence band for low Fe doping [e.g. 22]Generally, our results are only in qualitative agreement with the calculations available for low doping levels of iron in STO [9,10,11]. The calculations indicate to presence of isolated contribution from Fe mostly in the region of the gap close to the bottom of the conduction band. One has to remember that calculations are based on the presence of $Fe^{4+}$ states for Fe substituting Ti. However, da Silva et al. [9] and Blokhin et al [12] found in Fe K edge XANES indication to the lower oxidation state. Unfortunately, there is no data on PDOS calculated for Fe in the lower oxidation state than 4+. Moreover, calculations are performed in the ground state and final state effects inseparably related to photoemission are not included. On the other hand our results can be compared to the resonant photoemission data for Fe oxides published by Lad and Henrich [23]. They analyzed resonantly enhanced photoemission through the Fe 3p-3d excitation threshold for $Fe_xO$, $Fe_2O_3$ and $Fe_3O_4$, so for the compounds with primarily 2+, pure 3+ and mixed 3+ and 2+ Fe valence. Interestingly, the structure of the Fe 3d states for $Fe_2O_3$ has several features similar to our PDOS for $Fe^{3+}$, like the clear peak at about 3 eV, broad feature in the region 5-8 eV and another one at 12-13 eV. It can be understood when taking into account the structure of $Fe_2O_3$ where $Fe^{3+}$ cation is surrounded by a distorted octahedron of $O^{2-}$ ions. The mostly divalent Fe in $Fe_xO$ shows the main features at 1.2 eV and 4.1 eV and again this tendency is reflected in our Fe PDOS ascribed to $Fe^{2+}$ states. Also for this compound most Fe cations have the octahedral coordination.

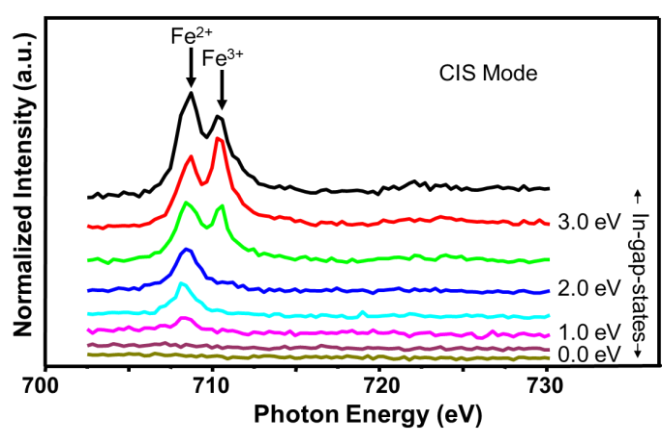

Fig. 5. Photoelectron intensity obtained for 2 % Fe doped STO film in the CIS mode presenting the photon energy dependence of the in-gap states.

For better understanding of the defect chemistry, relation between Fe valence state and oxygen deficit the samples were preheated in ultra-high vacuum at 630°C for 15 min. As it was suggested in the previous paper [15] such annealing leads to the increase of the oxygen vacancies in the surface region, and consequently to relative increase of the $Fe^{2+}$ species which are connected to oxygen vacancies. However, as can be seen from the Fig. 4 the effect of annealing the films at 630°C is very weak. A slight increase of the density of states in the gap is visible for the resonance related to $Fe^{2+}$ ions for both samples doped with 2 and 5 % Fe. The $Fe^{3+}$ resonance shows a weak effect of annealing at higher temperature visible as a slight decrease of the main peak at 2.5 eV. Such behaviour can be understood as the effect of creation the additional oxygen vacancies in the surface region leading to the increase of the $Fe^{2+}/Fe^{3+}$ ratio.

Fe doping is nominally connected to substitution of Ti ions in the $TiO_6$ octahedra. As it was discussed in our earlier paper the detected valence state of iron is 2+ or 3+. The main question is then related to the number of oxygen vacancies associated to Fe ions centering the octahedra. Calculations by Alexandrov et al [11] for $Fe^{4+}$ indicate to partly covalent bonding between Fe and O atoms. On the other hand Blokhin et al [12] showed that oxygen vacancies ($V_O$) are located in the first coordination of $Fe^{3+}$ ions in the cathodic region of the electrocolored Fe-doped STO. It is thus not clear if the $Fe^{2+}$ state can be associated to the double oxygen vacancy within the octahedron. A possible explanation can be related to agglomeration of the Fe ions and oxygen vacancies forming extended defect complexes [15].

Zhou et al. calculated DOS for Fe doped STO at two concentrations of 5 and 12.5 % and Fe in the Ti and Sr sites [10]. They found a significant contribution of Fe originating states to the in-gap region. Despite the wrong value of the energy gap coming from the used method (GGA), the relatively broad band of Fe states spreading from the top of the valence band to about 1 eV below the bottom of conduction band was obtained for Fe at Ti site. For Fe at Sr site two sharp peaks of PDOS within the gap were found. Our results agree qualitatively with the case of 5% Fe at the Ti site. It is worth to mention also qualitative agreement with the results of Rothschild et al. [21] obtained with use of different methods.

To check the possible correlation between titanium and iron states, the photoemission studies of titanium electronic states were performed. For this reason we have measured the XAS spectra for all samples at the energy range assigned to the titanium $L_{2,3}$ absorption edge. An example of the spectrum for STO:2% Fe is presented in Figure 6. For all investigated samples the XAS spectrum had the same shape. It consists from four well separated peaks at 458.4eV, 460.8eV, 463.8eV, 466.1eV. All spectra showed also two small pre-peaks at energies below

458.4eV. The first of two main peaks, marked as "A" and "B" can be assigned to the $L_3$ edge ($t_{2g}$ and $e_g$ orbitals respectively) and the second two ("C" and "D") to the $L_2$ edge ($t_{2g}$ and $e_g$ respectively). The structure of the obtained XAS spectra is in accordance with the published data for STO.

To calculate PDOS of titanium the "B" peak (460.8 eV) was chosen to perform the "on resonance" measurements of the valence band electronic states. For the "off-resonance" measurements the photon energy was chosen to 452.8 eV. The selection of the resonance energy was caused by the presence of the feature coming from the second order light visible as an oblique line in the lower right corner of the photoemission map (Fig. 7). This feature disturbs observation of very weak effect in the energy gap region. The photoemission map shows the valence band photoemission of the STO:1% Fe film for photon energy range of the $L_{2,3}$ Ti absorption edge. The films with higher Fe content showed an analogous behaviour with slight differences observed within the energy gap region. Contrary to the photoemission map of the Fe resonance there is no damping of the intensity from the $L_2$ edge, caused by a lack of spin polarization of Ti states.

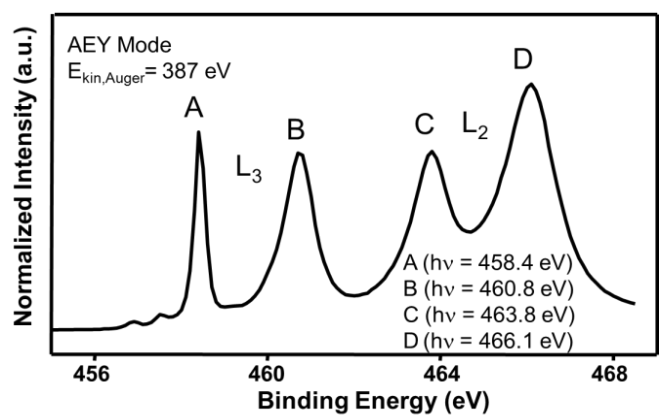

Fig. 6. Ti L edge x-ray absorption spectrum obtained in the Auger electron yield mode for the STO film doped with 2 % Fe.

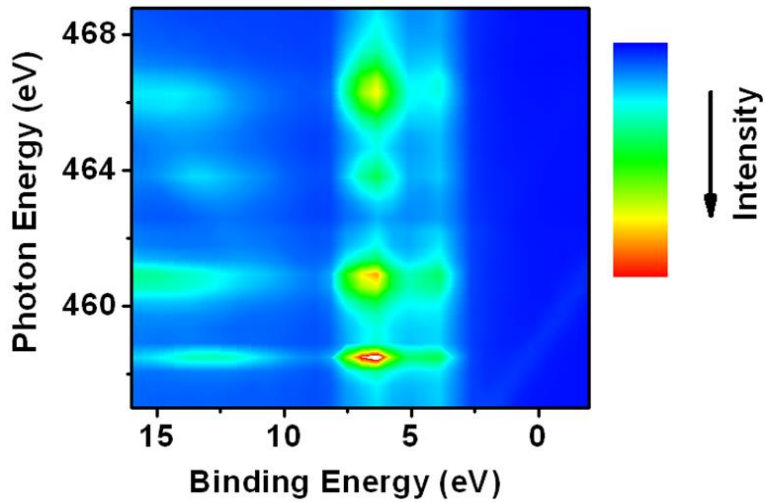

Fig. 7. Photoemission intensity maps in the valence band region obtained for the 1% Fe doped STO film for photon energies within the Ti 2p-3d resonance.

The main contribution of titanium states to the valence band is visible in the range 3-8 eV. This behaviour was observed also in STO bulk crystals [24,25]. Additionally, we found the states originating from Ti in the energy region 10-16 eV. Due to the observed photon energy dependence one can attribute them to the Auger transitions [26]. Analysis of the photoemission map does not indicate to a significant contribution of Ti states to the energy gap region. To observe a possible weak intensity in that region we calculated Ti PDOS in similar way as in the case of iron.

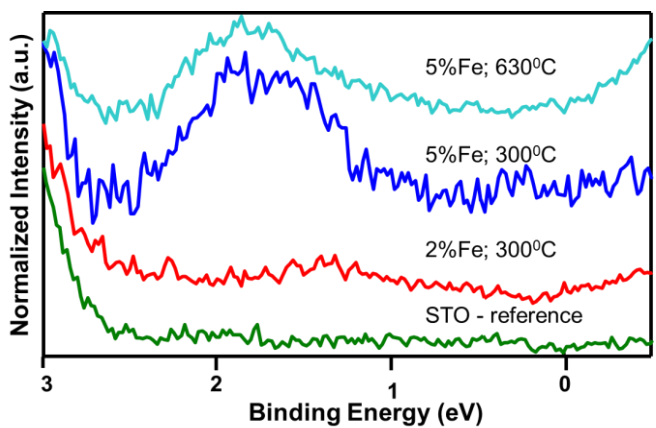

Fig. 8 The in-gap region of the valence band being a part of the Ti PDOS obtained by subtracting the photoemission spectra recorded at photon energies of 460.8 eV (on resonance) and 452.8 eV (off resonance). The spectra were normalized to the intensity at the main contribution to the valence band at 6.5 eV, not shown. Increasing intensity above the Fermi level comes from the second order light.

In Figure 8 the in-gap PDOS for titanium is presented for selected films and two temperatures of annealing. The increasing intensity for energies above the Fermi level is connected to the mentioned second order light from the monochromator. The spectra were normalized to the intensity of the main peak in the valence band situated at the energy of 6.5 eV. It is known that $Ti^{3+}$ states give a contribution to the in-gap region in reduced or electron doped STO [16,27]. The appearance of the $Ti^{3+}$ states can be related to the presence of oxygen vacancy and/or neighbouring electron donor. Ishida et al. [16] distinguished the coherent states which were situated at the Fermi level and incoherent ones which were ascribed to the local $Ti^{3+}$ states. The contribution in the energy range 1-2 eV has been ascribed to $Ti^{3+}$ incoherent states [16]. It is worth to mention that we found no reduced Ti states within the photoemission Ti 2p multiplets. However, the resonant photoemission clearly shows a Ti contribution to the in-gap states which is higher for the Fe doped films. Fe doping clearly increases this contribution although also for those films we found no reduced Ti states within the Ti 2p photoemission. The reason is that the resonant photoemission is able to detect much lower concentrations of these states due to the strong enhancement of emission from the Ti 3d band.

Calculations of DOS by Zhou et al. [10] indicated to a presence of some Ti 3d states in the gap for 5% Fe doped STO which was the results of hybridization with the Fe 3d states. Similar effect was found in calculations of Alexandrov et al [11]. This effect can be responsible for the observed in our studies increasing Ti PDOS with Fe doping level. Another explanation can be assigned to the increased amount of the oxygen vacancies being the result of Fe doping as it was discussed by Koehl et al. [15]. Thus, one could attribute the observed intensity at energies of about 1.5 – 1.8 eV to the complexes $Ti^{3+}$-$V_O$-Fe. The valence state of iron in such isolated complex would be rather 2+. It agrees with the hypothesis of hybridization between the Ti 3d states and Fe 3d ones. It is not clear why maximum of the Ti contribution to the in-gap states is shifting with Fe doping level.

Calculations taking into account the presence of divalent and trivalent Fe states and assuming localized complexes with oxygen vacancies are necessary to have fuller image of the electronic structure of doped STO.

**Conclusions**

The electronic structure of Fe doped epitaxial films of STO has been studied with the use of resonant photoemission. Based on our earlier analysis of the XAS studies, the Fe contribution to the valence band was divided into two parts related to the Fe valence states 2+ and 3+. These valence states result in different contributions to the valence band, especially within the in-gap region. The Fe3+ states form a sharp peak just above the top of the valence band while Fe2+ contribute as a rather broad feature situated in the binding energy range below 0.5 eV. No states at the Fermi level were observed. Resonant photoemission at the Ti 2p-3d edge showed an in-gap contribution from the Ti states increasing with Fe doping level. The effect can be attributed to the hybridization between the Ti and Fe originating electronic states to the valence band and with the presence of Ti states associated to Fe ions forming complexes with oxygen vacancies. The obtained partial DOS's are in qualitative agreement with the calculations for Fe doping level in STO below 10 % showing some states within the energy gap. However, calculations including defect complexes containing Fe should be performed in order to clarify the overcompensation of Fe by oxygen observed in thin films.

**Acknowledgements**

This work was supported by the NCBiR as well as the NRW Bank within the project ERA-NET-MATERA/3/2009. The work was furthermore partially funded by the Deutsche Forschungsgemeinschaft SFB 917 ("Nanoswitches"). The assistance of Karina Shulte at the I311 beamline at Max-Lab is greatly appreciated.

**References**

[1] Y. Iwasa, M. Kawasaki, B. Keimer, N. Nagaosa, Y. Tokura, and H. Y. Hwang, Nature Mater. **11**, 103 (2012).

[2] K. Szot, W. Speier, G. Bihlmayer, and R. Waser, Nature Mater. **5**, 312 (2006).

[3] F. J. Morin, and J.R. Oliver, Phys. Rev. B 8, 5847 (1973).

[4] R. Waser, T. Bieger, and J. Maier, Solid State Commun. **76**, 1077 (1984).

[5] R. Merkle, and J. Maier, Angew. Chem., Int. Ed. **47**, 3874 (2008).

[6] R. Moos, and K. H. Härdtl, J. Am. Cerami. Soc. **80**, 2549 (1997).

[7] T. Takeda, Y. Yamaguchi, and H. Watanabe, J. Phys. Soc. Jpn. **33**, 967 (1972).

[8] L. F. da Silva, M. I. B. Bernardi, L. J. Q. Maia, G. J. M. Frigo, and V. R. Mastelaro, J. Therm. Anal. Calorim. **97**, 173 (2009).

[9] L. F. Da Silva, J.-C. M'Peko, J. Andrés, A. Beltrán, L. Gracia, M. I. B. Bernardi, A. Mesquita, E. Antonelli, M. L. Moreira, and V. R. Mastelaro, J. Phys. Chem. C **118**, 4930 (2014).

[10] X. Zhou, J. Shi and C. Li, J. Phys. Chem. C **115**, 8305 (2011).

[11] V. Alexandrov, J. Maier, and R. A. Evarestov, Phys. Rev. B **77**, 075111 (2008).

[12] E. Blokhin, E. Kotomin, A. Kuzmin, J. Purans, R. Evarestov and J. Maier, Appl. Phys. Lett. **102**, 112913 (2013).

[13] Ch. Lenser, A. Kuzmin, J. Purans, A. Kalinko, R. Waser, and R. Dittmann, J. Appl. Phys. **111**, 076101 (2012).

[14] J. Szade, K. Szot, M. Kulpa, J. Kubacki, Ch. Lenser, R. Dittmann, and R. Waser, Phase Transitions **84**, 489 (2011).

[15] A. Koehl, D. Kajewski, J. Kubacki, C. Lenser, R. Dittmann, P. Meuffels, K. Szot, R. Waser, and J. Szade, Phys. Chem. Chem. Phys. **15**, 8311 (2013).

[16] Y. Ishida, R. Eguchi, M. Matsunami, K. Horiba, M. Taguchi, A. Chainani, Y. Senba, H. Ohashi, H. Ohta, and S. Shin, Phys. Rev. Lett. **100**, 056401 (2008).

[17] A. F. Santander-Syro, O. Copie, T. Kondo, F. Fortuna, S. Pailhès, R. Weht, X. G. Qiu, F. Bertran, A. Nicolaou, A. Taleb-Ibrahimi, P. Le Fèvre, G. Herranz, M. Bibes, N. Reyren, Y. Apertet, P. Lecoeur, A. Barthélémy, and M. J. Rozenberg, Nature **469**, 189 (2011).

[18] Y. J. Chang, A. Bostwick, Y. S. Kim, K. Horn, and E. Rotenberg, Phys. Rev. B **81**, 235109 (2010).

[19] Y. Adachi, S. Kohiki, K. Wagatsuma, and M. Oku, J. Appl. Phys. **84**, 2123 (1998).

[20] R. Waser, T. Bieger, J. Maier, Solid State Commun. 76 1077 (1990)

[21] A. Rothschild, W. Menesklou, H.L. Tuller and E. Ivers-Tiffe´e, Chem. Mater. 18, 3651, (2006)

[22] I. Denk, W. Münch, J. Maier, J. Am. Ceram. Soc. 78, 3265 (1995)

[23] R. J. Lad, and V. E. Henrich, Phys. Rev. B **39**, 13478 (1988).

[24] Y. Haruyama, S. Kodaira, Y. Aiura, H. Bando, Y. Nishihara, T. Maruyama, Y. Sakisaka, and H. Kato, Phys. Rev. B **53**, 8032 (1996).

[25] M. Takizawa, K. Maekawa, H. Wadati, T. Yoshida, A. Fujimori, H. Kumigashira, and M. Oshima, Phys. Rev. B **79**, 113103 (2009).

[26] T. Kaurila, R. Uhrberg, and J. Väyrynen, J. Electron. Spectrosc. Relat. Phenom. **88**, 399 (1998).

[27] Y. Aiura, I Hase, H. Bando, T. Yasue, T. Saitoh, and D.S. Dessau, Surf. Sci. **515**, 61 (2002).